\documentclass[aps,pra,preprint,groupedaddress]{revtex4-2}
\usepackage[top=2.5cm,bottom=2.5cm,right=3cm,left=3cm]{geometry}
\usepackage[utf8]{inputenc}
\usepackage{braket}
\usepackage{amsmath}
\usepackage{bbm}
\usepackage{appendix}
\usepackage{amsfonts}
\usepackage{comment}
\usepackage{cancel}
\usepackage{bbold}
\usepackage{graphicx}
\usepackage{float}
\usepackage{MnSymbol}
\usepackage{mathtools}
\usepackage{soul}
\usepackage{tabularx}
\usepackage{array}
\usepackage{accents}
\usepackage{bm}
\usepackage{placeins}
\usepackage{mathrsfs}

\newcolumntype{Y}{>{\centering\arraybackslash}X}

\usepackage[colorlinks, linkcolor=blue]{hyperref}
\usepackage[nameinlink, capitalise]{cleveref}

\newcounter{definition}
\newtheorem{Definition}[definition]{Definition}

\newcounter{theorem}
\newcounter{proposition}
\newtheorem{Proposition}[proposition]{Proposition}

\newcommand{\domark}{%
	\vbox to 0pt{
		\kern-\dp\strutbox
		\hbox{\smash{\llap{*\kern1em}}}
		\vss
	}%
}

\begin{document}

	\title{Dynamics in an emergent quantum-like state space generated by a nonlinear classical network}
	\author{Gregory D. Scholes}
	\address{Department of Chemistry, Princeton University, Princeton, NJ 08544, USA}
	\email{gscholes@princeton.edu}

	\date{\today}

	\begin{abstract}
	In recent work we have developed a theory for classical systems that mimic the state space of quantum systems. We call the states `quantum-like' (QL) because they show many properties similar to those of a quantum system. Here we study the QL states of a large classical system comprising a network of coupled phase oscillators. In particular, we seek to understand how the nonlinearity of the classical system manifests in the linear representation of the state space. To that end, we study how a classical network is viewed from the perspective of the time evolution of its emergent state. We investigate how mixedness of the QL states diminish or increase as the underlying classical system synchronizes or de-synchronizes respectively.  In the limit of a very strongly phase-locked classical network---that is, where couplings between phase oscillators are very large---the state space evolves according to unitary dynamics. When the system is set up so that the classical phase oscillators synchronize with time, we find that the purity measure of the QL states correspondingly increases. Whereas, in cases where the coupling between oscillators is too small to allow the network to synchronize, the classical variables act like a kind of environment that promotes dephasing of initially-prepared superpositions.  
	\end{abstract}

	\maketitle
    \newpage


\section{Introduction}

In recent work we have described how a classical system can be designed to produce a state space that has many properties similar to a quantum state space\cite{ScholesQLstates, QLproducts}. These classical systems, that we call quantum-like (QL), can be identified with networks that have a particular structure of phase relationships. Here we study the time evolution of the QL states that are associated with a QL network that evolves by nonlinear dynamics. In particular, how does the nonlinearity of the classical system manifest in the linear representation of the state space?

\begin{figure*}
	\includegraphics[width=15 cm]{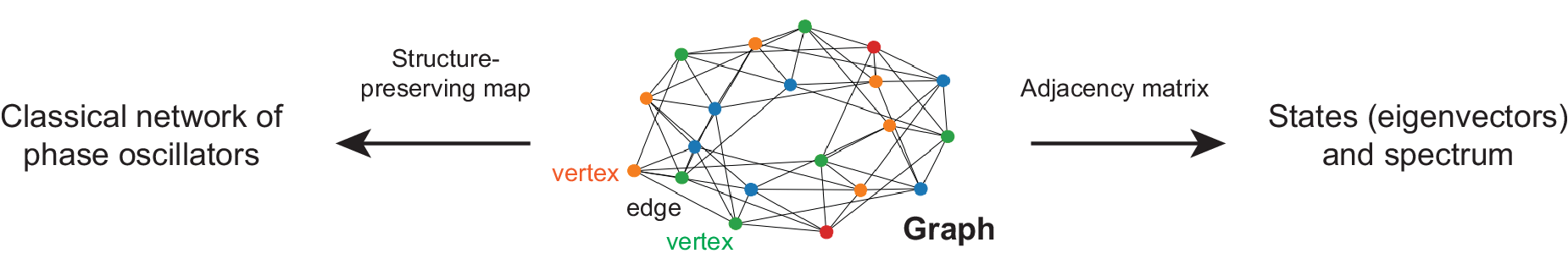}
	\caption{Picture of a graph with two vertices and the edge connecting them labeled. The adjacency matrix defined by the graph gives a set of states (the `state space') and spectrum. A map that preserves the structure of the graph in terms of its edge topology defines the structure of a corresponding classical network, for example a phase oscillator network. The map we use in the present work assigns a phase oscillator to each vertex of the graph and and a coupling between oscillators to each edge.}
	\label{figGraph}
\end{figure*}

In our work on QL systems, we use graphs to represent the structure. Graphs depict the way the basic building blocks (indicated by the vertices of a graph) are connected. A graph $G(n,m)$, or simply $G$, comprises $n$ vertices and a set of $m$ edges that connect pairs of vertices, Fig. \ref{figGraph}. Any graph is associated with an adjacency matrix $A$ that specifies the connections between vertices, called edges. The adjacency matrix is indexed by the vertices and contains an entry at $a_{ij}$ if the vertex $i$ is linked by en edge to vertex $j$, and a corresponding complex conjugate of that entry at $a_{ji}$ as long as the graph is undirected. The entry $a_{ij}$ is usually taken to be 1, but it could be any value. For more specialized background on graph theory see \cite{Diestel,Janson2000, Bollobas2001}. The adjacency matrix is similar to a Hamiltonian matrix if we consider the edges to take values of coupling matrix elements. Similarly to the Hamiltonian matrix, $A$ can be diagonalized to give eigenvectors and a spectrum for the graph.

\begin{figure*}
	\includegraphics[width=10 cm]{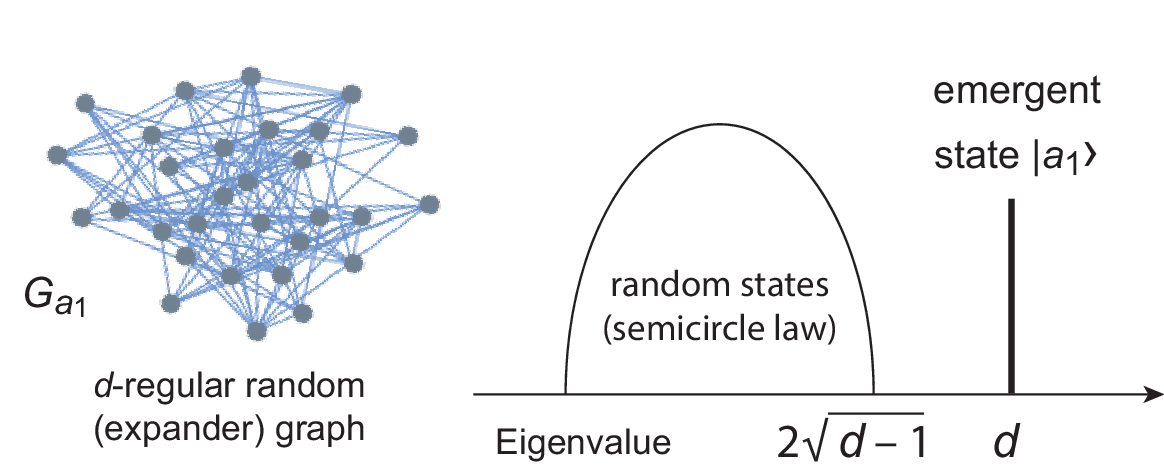}
	\caption{As a starting point for QL graphs we need a source of states. We use the emergent state $|a_1\rangle$ from an expander graph, $G_{a1}$, as a state. This state is guaranteed to be separated in the spectrum from all the other states (that we ignore), which is why it is called an \emph{emergent state}. The other states in the spectrum, labeled the random states, have a semi-circular density-of-states like the spectrum of a random matrix. The emergent state is robust because of this separation from other states in the spectrum.}
	\label{figEmergent}
\end{figure*}

The graphs considered in the present work contain many ($n$) vertices, so the adjacency matrix gives $n$ eigenvectors (that we refer to  as states). However, by considering particular kinds of graphs (described in the next section), we can focus on a single so-called \emph{emergent} state. An emergent state has an eigenvalue that is well-separated from the rest of the spectrum. The particular emergent state we use to build the QL states from graphs is the `trivial' eigenvector of an expander graph, which is the constant vector (all coefficients are equal), Fig. \ref{figEmergent}. Physically, the fact that the emergent eigenvector is the constant vector indicates that the graph's stationary distribution for a random walk is uniform. Consistent with the mathematical literature, we show the emergent state to have the largest eigenvalue in the spectrum. However, when we study phase oscillator networks, the sign of the couplings are reversed, so the emergent state will be the ground state of the system.

A graph can also define the layout of a network\cite{Barabasi}. Networks have been widely studied, in particular with regard to the phenomenon of synchronization. The literature on synchronization phenomena is vast\cite{StrogatzBook, Strogatz2000, Strogatz1993,Acebron2005,daFonseca2018, Ha2016, Gherardini2018, Sync1}. The idea of collective motion is well known to us from everyday experiences, and provides a good concrete example of physical synchronization\cite{Collectivemotion, Flocks}. There are numerous examples, from circadian rhythms to cell cycles to neural networks to ensembles of atoms\cite{circadian, rhythmicphysiol, bacteriallysis, SyncSpiking,  corticalnetworks, brain6, atomensembles}. The basis for models of synchronization is reviewed in Appendix A. The structure of the network plays an important role in determining synchronization the propensity of a system to synchronize\cite{Townsend2023, Townsend1, Townsend2, Townsend3}. If we represent the QL graph as a network (in the present work, a network of oscillators), what is the difference between the graph and the classical network, because they appear to look the same? The answer is that they can be the same, but this is not required. For example, the classical network could have more or less vertices than the graph, as long as it has the same structure (edge topology). 

In this report we study synchronization in a phase oscillator network that has the structure of graphs that produce QL states. We can use a widely-known model for synchronization phenomena as an example of a classical network that undergoes nonlinear dynamics. The main question to be addressed in the present work is how the nonlinear classical dynamics of the network show up in the time evolution of the QL states, which are a linear representation of the network because they come from the graph's adjacency matrix. Using numerical studies, we study how the QL states associated with the linearized system---the graph's adjacency matrix---evolve.  

In what limiting case is the time evolution of the QL states unitary? Outside of that limit, what is the physical interpretation of the dynamics of the QL states? Thus, we explore the way synchronization and de-synchronization of the classical system translates to time evolution of the QL state space.  Synchronization occurs when the coupling between phase oscillators is sufficiently large that the phases become locked in step; a phenomenon that is extensively studied. When the network is synchronized, we have concomitant emergent QL states. Dephasing of the QL states results from the opposite limit, where couplings are small and the phase oscillators evolve according to their natural frequencies and the network loses phase coherence. As a result, emergent QL states are absent.

This paper is structured as follows. In the Sec. II, the QL graphs and their states are introduced. Then in Sec. III, details are given for the map to the classical network that will be studied using numerical calculations in Sec. IV. Finally, the broader implications of the study are summarized in Sec. V.

\section{Construction of quantum-like states from graphs}

The starting point for graphs that produce QL states is a graph that is a good source of states. By `good', what we mean is that we want a really large and complex graph---imagine part of a neural network in a brain---to have a spectrum that shows one special state that is isolated from all the other many (possibly infinite even) states. That is, we need an emergent state. An example of a graph that will not work as a reliable source of states is one that represents a system of vertices connected only by nearest-neighbor edges, because as the system gets larger, its spectrum gets more congested\cite{Excitonics2024}. In contrast, we need a graph that \emph{guarantees} one state is separated from the rest. We therefore focus on special graphs, called expander graphs, that have a spectrum comprising a distinguished emergent state---a state that is clearly separated from other states in the spectrum. The emergent state of an expander graph, Fig. \ref{figEmergent}, is the resource we use to construct the QL state space. 

Expander graphs\cite{Sarnak2004, expandersguide, Lubotzky, Expanders, Expanders2, Alon1986, Tao-expanders} are highly connected graphs that are optimal for random walks and communications networks. The physical concept underpinning an expander is that the edges are scale-free, so that no matter how we lay out the vertices, edges connect vertices at all length scales. A subset of the $d$-regular graphs have this property where edges connect vertices over many distance scales (distance defined with respect to the graph), and are prototypical expander graphs. However, it is not so easy to pinpoint which $d$-regular graphs are expander graphs. Nevertheless, $d$-regular \emph{random} graphs are asymptotically almost certainly good expanders\cite{Friedman1991, Miller2008}. These are the graphs we use for the QL bit subgraphs. See Ref. \cite{Excitonics2024} for background. 

\begin{figure}
	\includegraphics[width=15 cm]{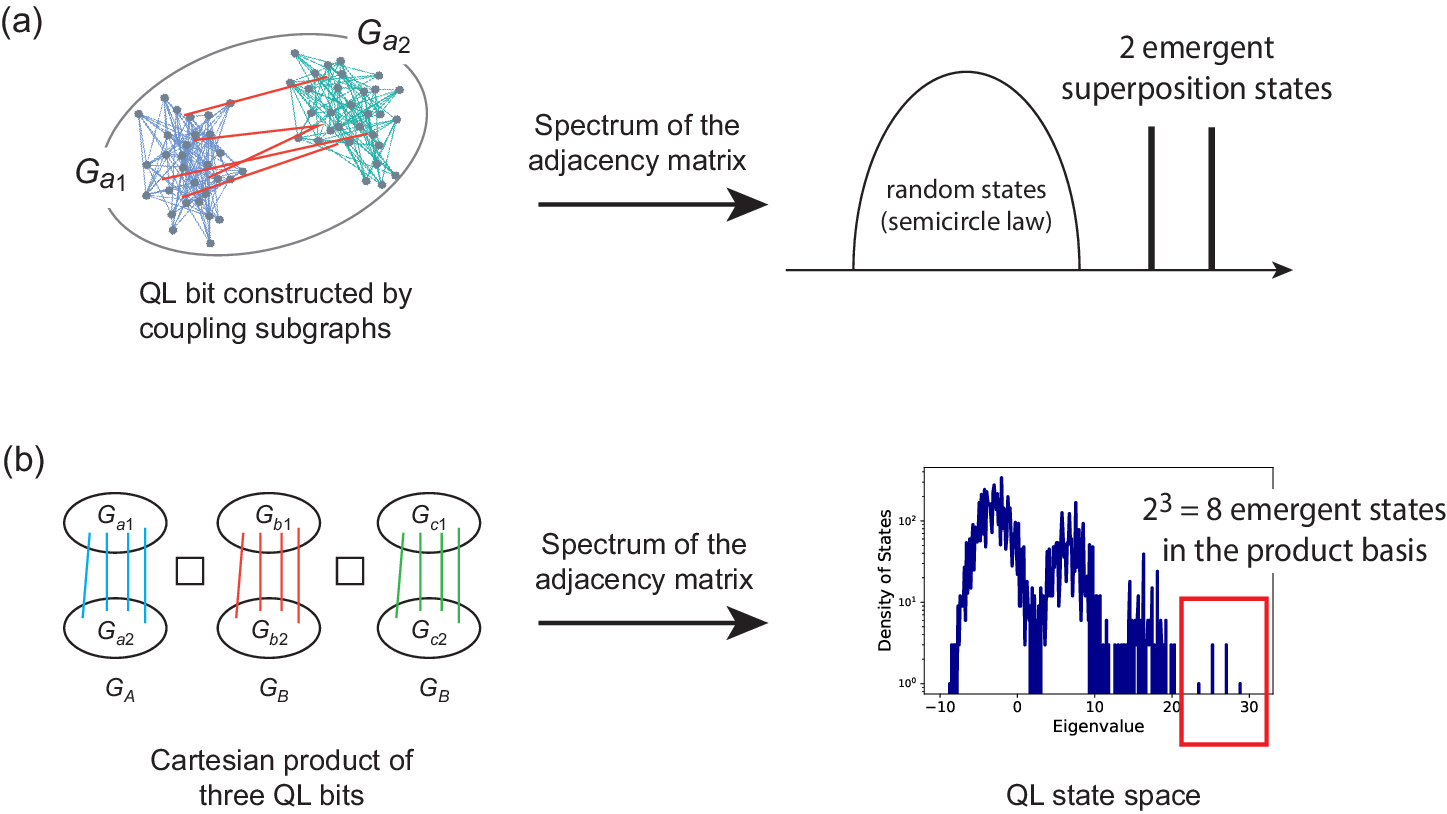}
	\caption{(a) To produce a system with two emergent states, a QL bit, we ‘hybridize’ two expander graphs by connecting them with additional edges. Now we have two emergent states that are superpositions of the states of the two subgraphs. (b) From this foundation we generate arbitrary states in the tensor-product basis. We do that by combining the QL bit graphs using an operation called the Cartesian graph product. The states of a product graph are tensor products of states of the graphs combined in the product. }
	\label{figQLsummary}
\end{figure}   

Two expander graphs are linked together by connecting edges to produce a two-state system that we call the QL bit\cite{ScholesQLstates}, Fig. \ref{figQLsummary}a. We are focusing on the emergent state of each subgraph and their superposition in the QL bit states. This indicates that we should change our basis from that defined by all the vertices of each graph to a basis defined by the QL bit subgraphs. We now show how to perform that basis change. We enumerate the basis of $G_A$ (with $n$ vertices) as basis states $\{ u_1, u_2, \dots, u_n\}$ for subgraph $G_{a1}$, and $\{x_1, x_2, \dots, x_k \}$ for subgraph $G_{a2}$ (with $k$ vertices). Taken together, we have a complete orthonormal basis in $\mathbbm{C}^{(n+k)}$. Now, an arbitrary emergent state, $W_i$, of the QL bit graph will be written in terms of this basis as:
\begin{equation}
	W_i = c_1u_1 + \dots + c_nu_n + c_{n+1}x_1 + \dots + c_{n+k}x_k
\end{equation}
where the $c_l$ are complex coefficients. To contract the basis to be defined in terms of subgraphs we identify the basis states associated with each subgraph using
\begin{eqnarray}
	J_{a1} = \{ \underbrace{1, 1, \dots, 1,}_{\text{\emph{n} times}} \underbrace{0, 0, \dots, 0}_{\text{\emph{k} times}}  \}/\sqrt{n} \\
	J_{a2} = \{ \underbrace{0, 0, \dots, 0,}_{\text{\emph{n} times}} \underbrace{1, 1, \dots, 1}_{\text{\emph{k} times}}  \}/\sqrt{k}.
\end{eqnarray}
We can  then resolve the coefficients for the effective two-states of the QL bit in terms of the basis associated to the subgraphs $G_{a1}$ and $G_{a2}$, $|a_1 \rangle$ and $|a_2 \rangle$ using the inner products 
\begin{eqnarray}
	\alpha_i = \langle J_{a1}, W_i \rangle \\
	\beta_i = \langle J_{a2}, W_i \rangle 
\end{eqnarray}
where $i$ labels the state of the QL bit, ordered by eigenvalue. We thus obtain the effective states of the QL bit as
\begin{equation}
	W_i^{2 \times 2} = \alpha_i |a_1 \rangle + \beta_i |a_2 \rangle .
\end{equation}

Now we generalize our model for the QL bit graph to allow the edges to take values, or a bias, of any number on the unit circle in the complex plane. To illustrate this, consider the eigenvectors for the effective adjacency matrix for the QL bit graph comprising two $d$-regular subgraphs, where the edges within each subgraph have a bias of 1, and the edges that couple the two subgraphs have a bias of $e^{i\gamma}$. Then:
\begin{equation}
	A_{2 \times 2} = 
	\begin{bmatrix}
		d & e^{i\gamma} \\
		e^{-i\gamma} & d 
	\end{bmatrix} ,
\end{equation}
which gives an eigenvector for the emergent state $(1, e^{-i\gamma})$ (neglecting normalization). Note that we have omitted an overall phase of these states. When that phase is accounted for, then we see that the emergent state is any vector in $\mathbb{C}^2$. 

We can continuously rotate this connecting edge bias to generate states (when the system is perfectly synchronized) that emulate, for example, the states of a system of spins or polarization states of photons. The notable emergent states isomorphic to polarization states of light are collected in Table 1. 

 \begin{table}[b]
 	\caption{\label{tab:table1}
 		An example of a set of QL graph edge biases and the corresponding QL emergent states (not normalized).
 	}
 	\begin{ruledtabular}
 		\begin{tabular}{cccc}
 			\textrm{$a_1$ bias}&
 			\textrm{$a_2$ bias}&
 			\textrm{Connecting bias}&
 			\textrm{State}\\
 			\colrule
 			+1 & +1 & +1 & $|a_2\rangle + |a_1\rangle $ \\
 			+1 & +1 & -1 & $|a_2\rangle - |a_1\rangle $ \\
 			+1 & +1 & -$i$ & $|a_2\rangle + i|a_1\rangle $ \\
 			+1 & +1 & $i$ & $|a_2\rangle - i|a_1\rangle $ \\
 		\end{tabular}
 	\end{ruledtabular}
 \end{table}

\begin{figure}
	\includegraphics[width=6 cm]{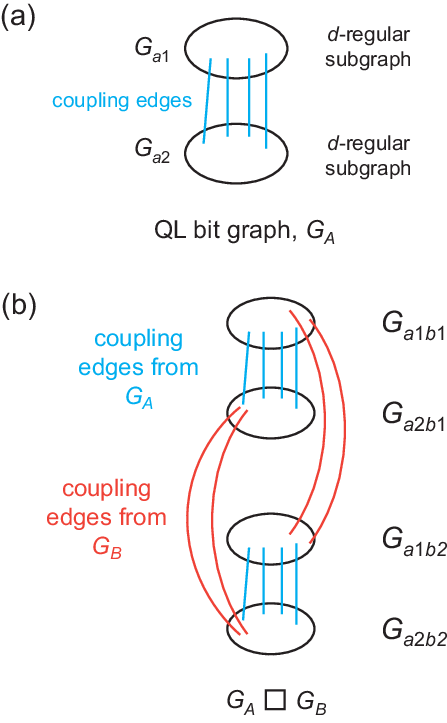}
	\caption{(a) Schematic drawing of the a QL bit graph. (b) Structure of the Cartesian product. Notice how the four subgraphs define the basis as a product of the basis from each QL bit graph. }
	\label{figQL2}
\end{figure}

The QL bit graphs, in turn, are combined using a graph operation---the Cartesian product of graphs---to generate a new, exponentially larger, graph, Fig. 3b. Background on graph products is given in Appendix B. The states of this product graph are separable linear combinations of tensor products of the states of the QL bits\cite{QLproducts}. In the present work we study networks that have the structure of the Cartesian product of two graphs, that is $G_A \Box G_B$. The structure of that graph is based on four subgraphs, and is shown schematically in Fig. \ref{figQL2}. 

The Cartesian product allows us to combine QL bit graphs so that a basis for eigenstates of a multi-QL bit system is a tensor product basis. In other words, the state space for each QL bit $i$ is defined in the Hilbert space $\mathcal{H}_i$, where the states are vectors such as $W_{2 \times 2} = \alpha|a_1\rangle + \beta|a_2\rangle$. A state space for $q$ QL bits is
\begin{equation}
	\mathcal{H} = \mathcal{H}_1 \otimes \mathcal{H}_2 \otimes \dots \otimes \mathcal{H}_q
\end{equation}

Our basis is then the set of $2^q$ states
\begin{equation}
	|a_i\rangle \otimes |b_j\rangle \otimes |c_k\rangle \otimes \dots ,
\end{equation}
where $i, j, k, \dots \in \{1, 2\}$ and states $a_i$ come from subgraphs of graph $G_A$, states $b_j$ come from subgraphs of graph $G_B$, and so on.

\section{Mapping QL state graphs to phase oscillator networks}

\subsection{Phase oscillator model for the classical network}

The Cartesian product of two QL bit graphs $G_A \Box G_B$ gives a new graph that generates the separable composite states of this system comprising two QL bits. In the following numerical work, we map this graph directly to a system of coupled phase oscillators. That is, each vertex of the graph becomes a phase oscillator with an assigned frequency and initial phase. Each edge of the graph indicates which phase oscillators are pairwise coupled.  Networks of \emph{phase oscillators} are especially appealing for providing physical insight into QL states because they focus on oscillator phase. The oscillators can be genuine oscillators, like pendulum clocks, or they can be concentrations of chemical reagents, or a range of other abstract concepts\cite{StrogatzBook}.

In the numerical simulations, each subgraph comprises $n_0$ vertices, giving a total of $N = 2n_0 \times 2n_0$ vertices in the product graph. For the calculations, each subgraph $G_{a1}$, $G_{a2}$, $G_{b1}$, $G_{b2}$ comprises 20 vertices and is $d$-regular with $d = 15$. Edges between subgraphs were added randomly, with probability 0.2. We calculate the full Cartesian product, yielding $2d$-regular subgraphs in $G_A \Box G_B$, with an eigenvalue of 30 for the emergent state when the subgraphs are uncoupled (that is, neglecting the eigenvalue shift labeled $\Delta$ in Ref. \cite{QLproducts}). We expect four emergent states, two with an eigenvalue of 30, one at $30 + \Delta$, and one at $30 - \Delta$. However, we focus on the emergent state with highest eigenvalue (in the mathematical notation). This will be the ground state in the numerical model.

In the phase oscillator model, each vertex can be thought of as an oscillator endowed with a frequency, that we treat in the rotating frame of the network of oscillators by its difference from the mean frequency $\epsilon_i$, and a phase offset $\phi_i$ for the oscillator at vertex $i$. We collect these terms as the time-dependent phase for each oscillator $\theta_i = \epsilon_it + \phi_i$, which is associated with the set of vertices. The oscillators are coupled according to the edges in the graph which, under appropriate conditions\cite{Strogatz2000, Acebron2005, daFonseca2018, Rodrigues2016, ScholesAbsSync}, allows the oscillators to synchronize after several periods of time. The phases associated to the vertices come into play as phase differences $e^{i(\theta_i -\theta_j)}$ multiplying the non-zero entries off-diagonal of the graph's adjacency matrix. This is accomplished by a suitable unitary transformation (specified below). The phases then evolve, according to the Kuramoto model\cite{daFonseca2018, Strogatz2000}, as:
\begin{equation}
	\dot{\theta}_i   = \epsilon_i - \frac{K}{N} \sum_{j=1}^{N} a_{ij} \sin(\theta_j - \theta_i),
\end{equation}
where $\epsilon_i $ is the frequency offset from the mean of the oscillator at node $i$, $\theta_i$ is the oscillator phase defined in terms of accumulated phase and an offset $\theta_i(t) = \epsilon_it + \phi_i$. $K$ is the coupling value, $a_{ij}$ are entries from the adjacency matrix of the graph. Notice that the coupling enters with a $-K/N$ prefactor, which makes the emergent state the ground state of the network and weights the edge by the factor $K$ renormalized by the number of oscillators in the network $N$. The nonlinearity comes from the coupling, which favors minimization of the phase differences.

The coupling $K$ is chosen to be 250 for the first set of calculations. Other choices of $K$ are reported in the subsequent results. Vertices are assigned a mean frequency of 100 in the same arbitrary units as $K$, and offset by frequencies randomly selected from a normal distribution with standard deviation 1 (in the same arbitrary units). Initial phases for the oscillators are randomly assigned from a von Mises distribution. 

The parameters in the calculations are selected using well-known properties of the Kuramoto model, Eq. 10. Whether or not a system synchronizes depends on the size of the coupling strength $K$ compared to the spread of frequencies $\nu_i$. This principle is clearly illustrated by a system of only two phase oscillators, see Sec. 8.6 of Ref. \cite{StrogatzNonlinDyn}. The phase portrait (i.e. $\theta_1$ is plotted versus $\theta_2$ on the torus) shows stable phase-locked solutions when $|\nu_1 - \nu_2| < 2K$. That is, the phase oscillators synchronize. Whereas, when $|\nu_1 - \nu_2| > 2K$ the flow resembles that of uncoupled oscillators (the coupling is insufficient to enable them to synchronize).  The $|\nu_1 - \nu_2| = 2K$ is a kind of dividing line, a saddle point  bifurcation. 

When there are many oscillators, the situation is more complex because we can find situations where some oscillators in the system synchronize with each other and others do not. That outcome will depend on the distribution of frequencies as well as the way the oscillators are coupled (i.e. the graph that defines the coupling matrix $[a_{ij}]$). Nevertheless, a rule of thumb for a system with \emph{all-to-all} coupling is that when $K$ in Eq. 10 sufficiently exceeds the standard deviation of oscillator frequencies ($\sigma_{\nu}$), the system will synchronize. For examples see Ref. \cite{Acebron2005, daFonseca2018}. The network studied here does not have all-to-all coupling, so we tested various couplings to find parameters where the system of phase oscillators will synchronize or will not maintain synchronization. 

In sum, whether or not a system of oscillators will perfectly synchronize depends not only on $K$ and $\sigma_{\nu}$, but also on the initial conditions. That is, the specific assignment of frequencies and initial phases to each oscillator in the network. For insight into this statement see, for example, Refs. \cite{Townsend1, Townsend2, Townsend2023, Taylor2012, Ling2019}. This sensitivity of the system to initial conditions emphasizes the nonlinearity of the network. 

\subsection{Mapping the classical phase oscillators to the QL states}

During the evolution of the system, the phase of each oscillator $\theta_i$ evolves nonlinearly. At selected time points of the numerical calculations of Eq. 10 the oscillators phases are mapped into the adjacency matrix of the QL state graph. To accomplish that, we transform the adjacency matrix of the graph as follows. The phases of the oscillators, which generally evolve nonlinearly, are the key outcome of the Kuramoto model.

At any time point in the evaluation of Eq. 10, each vertex $v_i$ in this large network is associated with a phase $\theta_i$. What we want to study is how the corresponding QL state---the one state with highest eigenvalue in the way we write the adjacency matrix---evolves. Notice that we write the Kuramoto equation so that the coupling term is preceded by a minus sign, so this state that has the largest eigenvalue for the adjacency matrix is the ground state of the classical model. To study the QL state under the Kuramoto dynamics, we need to modify the adjacency matrix so that it includes the evolving phases that are associated with the graph vertices. We do that by using a unitary transformation at each time step. 

The eigenvectors we have described in prior work, the QL states, are calculated for the normal adjacency matrices of the graphs (i.e. edges have values $\pm1$). They correspond to the states of perfectly globally-synchronized systems. More generally, however, each oscillator labeled $j$ is associated with a phase $\theta_j$. The adjacency matrix $A$ is transformed by these phases according to:
\begin{equation}
	A^{\prime} = \Phi^{-1}A\Phi ,
\end{equation}
with
\begin{equation}
	\Phi = 
	\begin{bmatrix}
		e^{i\theta_1} & 0 & 0 & \dots \\
		0 & e^{i\theta_2} & 0 & \dots \\
		0 & 0 & \ddots & 0 \\
		0 & \dots & 0 & e^{i\theta_{2n}} 
	\end{bmatrix}
\end{equation}

This unitary transformation of $A$ does not change the spectrum, but it does rotate the basis for the eigenstates. That basis change, in general, varies throughout the ensemble and therefore influences the mixedness of the state\cite{ScholesAbsSync}. We use the term `ensemble' to mean an average over initial phases to account for state preparation, as described in the following section.

The matrix $\Phi$ lists the phase of each oscillator in the system, associating each phase to a vertex representing an oscillator. Recall that this phase is the sum of accumulated phase associated with the frequency offset from the mean and a phase offset, $\theta_i(t) = \epsilon_it + \phi_i$. The transformation weights edges in the graph by phase \emph{differences}. Thus, for an edge connecting vertex $j$ to vertex $k$, the entry in the upper triangular part of the adjacency matrix transforms as $1 \rightarrow \exp(-i[\theta_j - \theta_k])$, with the corresponding complex conjugate value in the lower triangular part. The diagonal elements remain as entries of 0.

In the numerical examples shown below we obtain the phases from simulations using the Kuramoto model.  For any time point during the simulation, we use the oscillator phases to define $\Phi$, then we diagonalize $ \Phi^{-1}A\Phi $ to obtain the eigenvectors of the emergent states---now with complex coefficients.  This procedure introduces phase differences by biasing the edges between vertices $j$ and $k$ by $e^{i(\theta_k - \theta_j)}$. 

The phases in Kuramoto model evolve nonlinearly, but they transform the adjacency matrix through a unitary operator. By definition this preserves inner products, so the evolution of the state happens by a sequence of linear maps. We emphasize that statement below in more detail. The result is particular for the phase oscillator model and might not be generalizable. However, it is known that the nonlinear action of groups can be linearized by using a matrix representation of the group, which necessarily transforms according to linear algebra. This is more-or-less what we are trying to accomplish through the graph representation of the classical system.

\subsection{State preparation and averaging}

The evolution of the oscillator system, though sensitive to the specified initial conditions, is entirely deterministic. However, we use a particular averaging over the initial conditions that set up the numerical calculations as though cannot know the initial phases of the oscillators, only the variance of the distribution. These initial phase conditions are decided by a probabilistic state preparation. To account for such a state preparation, in the numerical studies we take an `ensemble' average over initial conditions of the phase oscillators which is intended to capture a scenario that we cannot know precisely the phases. It is not an average over a set of distinct QL systems.  

The von Mises distribution has either a very large circular standard  deviation, meaning the distribution is similar to a uniform distribution, or a small value, 0.001, meaning the distribution is extremely narrow. Results are averaged over either 100 or 500 realizations of this set-up. See ref \cite{ScholesAbsSync} for some background and examples that might be helpful. In the following subsection we discuss the Kuramoto model to provide context and justification for the choice of these parameters. 

The main motivation for the averaging over realizations of the initial phases is to account for uncertainty in preparation of the initial state. Therefore, we use the term `ensemble' to mean an average over initial conditions of oscillator phase. The initial phase for any oscillator refers to its natural frequency together with its phase offset, $\epsilon_i$ and $\phi_i$, respectively. We also checked calculations where we, additionally, average over an ensemble of graphs. 

We average the spectrum of the transformed adjacency matrix over all iterations of the simulation to give ensemble spectra. The spectra were computed for various time points, and therefore order parameters, as the system synchronized. Note that although the phases of the initial distribution of oscillators are randomly distributed, the phase differences contain embedded correlations---there are $N$ phases, but much more than $N$ phase differences. However, the phase differences are uncorrelated from one simulation to another, so mixedness comes from an essential ensemble averaging over simulations. In prior work mixedness as a function of synchronization was studied in some detail\cite{ScholesAbsSync}.

\subsection{Density matrix for the effective state of the composite system}

Recall the following orthonormal basis for a composite system in $\mathcal{H} = \mathcal{H}_A \otimes \mathcal{H}_B$ and equate it to the symbols we use for the analogous QL state basis:

\begin{eqnarray*}
	|0_A\rangle \otimes |0_B\rangle = |a_1\rangle |b_1\rangle \\
	|1_A\rangle \otimes |0_B\rangle = |a_2\rangle |b_1\rangle \\
	|0_A\rangle \otimes |1_B\rangle = |a_1\rangle  |b_2\rangle  \\
	|1_A\rangle \otimes |1_B\rangle = |a_2\rangle  |b_2\rangle .
\end{eqnarray*}

Any state in $\mathcal{H}$ can be written as a linear combination of these basis states. Furthermore, any separable (pure) state can decomposed as $W^{4 \times 4}= |\psi_A \rangle \otimes |\psi_B \rangle$, where $|\psi_A \rangle = W^{4 \times 4}_A = \alpha_A |a_1\rangle + \beta_A |a_2\rangle$ and $|\psi_B \rangle = W^{4 \times 4}_B = \alpha_B |b_1\rangle + \beta_B |b_2\rangle$. These separable states are generated by the Cartesian product of two QL bit graphs. As introduced earlier in the paper, here we focus on just one of the $W_i^{4 \times 4}$ states, that with largest eigenvalue. It corresponds to the `ground state' of the classical phase oscillator system (according to the way we write Eq. 10), and gives the instantaneous eigenvector of the dynamical network projected to the effective four-state model. This eigenvector does not vary with time once the phase oscillator system reaches equilibrium. 

To obtain this instantaneous eigenvector of the dynamical network projected to the effective four-state model, we take the oscillator phases and construct the unitary matrix $\Phi$, then transform the adjacency matrix of the graph according to Eq. 11.  Then we project this emergent state with highest eigenvalue onto the basis where each subgraph, labeled $a_1$, $a_2$, etc., maps to the corresponding single state $|a_1\rangle$, $|a_2\rangle$, etc. That gives all four of the $W_i^{4 \times 4}$, in terms of the complex coefficients $c^i_k(t)$ obtained from the projections.  The density matrix for the emergent state is calculated in this basis by statistically averaging over the pure states:
\begin{equation}
	W_i^{4 \times 4}(t) = c^i_1(t)|a_1\rangle |b_1\rangle + c^i_2(t)|a_2\rangle |b_1\rangle + c^i_3(t) |a_1\rangle  |b_2\rangle + c^i_4(t) |a_2\rangle  |b_2\rangle ,
\end{equation}
for each time point $t$ in a simulation. We obtain the coefficients $c^i_k(t)$ using a projection analogous to that described in Eqs. 1--6.

From the average over state preparations of $W_i^{4 \times 4}(t)$ we determine the density matrix $\rho(t)$ associated with the emergent state $i$ in the product basis, with entries
\begin{equation}
	\rho_{mn}(t) = \sum_{\text{prep}} c_m(t) c^*_n(t) 
\end{equation}
where $ \sum_{\text{prep}}$ means an average over state preparations. We report $\mathrm{Tr}(\rho)^2$ as a simple measure of mixedness of the state. Notice that the complex phases cancel to unity in the diagonal elements of $\rho_{mn}(t)$. On the other hand, the complex phases are retained in the off-diagonal elements and unless the system is synchronized, they interfere in the average and can thereby diminish or erase the off-diagonal elements, introducing state mixedness\cite{ScholesAbsSync}.

\section{Numerical results}

\subsection{Synchronization}

We now describe numerical results for a network of phase oscillators that are connected in a way prescribed by the product of two QL bits. The initial phase distribution was set to be close to the uniform distribution and the coupling is relatively strong, $K = 250$. These conditions allow us to observe synchronization of the oscillators. An average over initial phases is taken by iteration of the calculations.  Collective synchronization in an oscillator ensemble comprising $N$ oscillators is quantified by the order parameter, where $\text{Re}(x)$ means the real part of $x$:
\begin{equation}
	\text{order parameter} = \text{Re} \Big( \frac{1}{N} \sum_{i=1}^N e^{i\theta_i} \Big) .
\end{equation}
The line in Fig. \ref{figSync3}a shows the ensemble average order parameter predicted by propagating Eqs. 10 over 80 periods of the mean natural frequency of the oscillators. At early time the order parameter is small because the initial condition specifies a near uniform phase distribution. As time increases, the order parameter increases to approximately 1, showing that the nonlinear evolution of the phase oscillator system synchronizes the oscillator phases. 

\begin{figure*}
	\includegraphics[width=8.0 cm]{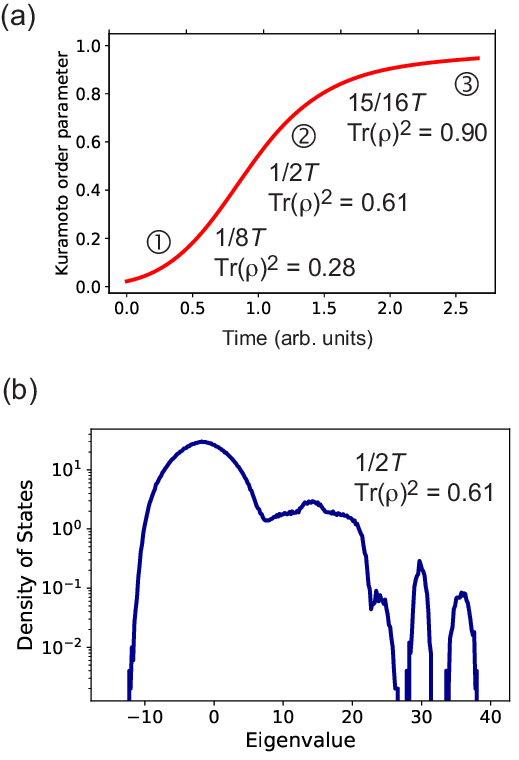}
	\caption{(a) Ensemble-averaged Kuramoto order parameter as a function of time predicted for the system of oscillators. The time points corresponding to the plots of spectra are indicated together with the purity calculated at that time point. The time points are indicated as a fraction of the total time of the simulation.  (b) Spectra calculated at one representative time point. }
	\label{figSync3}
\end{figure*}   

Concomitantly, the state associated with the greatest eigenvalue in the spectrum of the state space represented by the graph becomes less mixed. The state evolves in the effective state space according to the time dependence of the coefficients in Eq. 13. We highlight this result for the time points labeled numerically in Fig. \ref{figSync3}a. Corresponding spectra are all identical, as expected. An example is plotted in Fig. \ref{figSync3}b. The purity of the state with greatest eigenvalue is indicated together with the time point, written as a fraction of the total time of the simulation. For example, $\tfrac{1}{8}T$ equates to 10 periods of oscillation. Note that the purity for a fully mixed system of dimension 4, with a diagonal density matrix, is 0.25 (see Sec. 4d of Ref. \cite{ScholesQIS}). The initial purity of 0.25 means that the state is mixed because we have no information about the phases of the oscillators. As the phases synchronize, the variance of the phases diminishes, and the purity increases. The relationship between phase distribution and purity is discussed in Ref. \cite{ScholesAbsSync}.

The  evolution of the emergent state is generally not unitary. When the coupling $K$ for the classical system is very weak compared to the standard deviation of the frequency distribution, then the oscillators cannot synchronize and the QL states remain mixed. When the coupling is very large, then the oscillator system is locked in phase. In this regime, evolution in the QL state space can be unitary, as we show in Sec. IVc. That is, when the oscillators remain synchronized, state purity is preserved. Another regime is suggested by the Lohe model (see Appendix C). Here, all the oscillators are strongly synchronized \emph{within} each QL bit ($K$ is large within the subgraphs), but the coupling is weaker \emph{between} QL bits ($K$ is smaller for the connecting edges). Now the general model  for the single-excitation subspace specializes to the Lohe model. Single-excitation subspace means that the network graph carries the structure of the tensor \emph{sum} of QL bit graphs and QL bits are coupled to other QL bits. 

\begin{figure*}
	\includegraphics[width=15 cm]{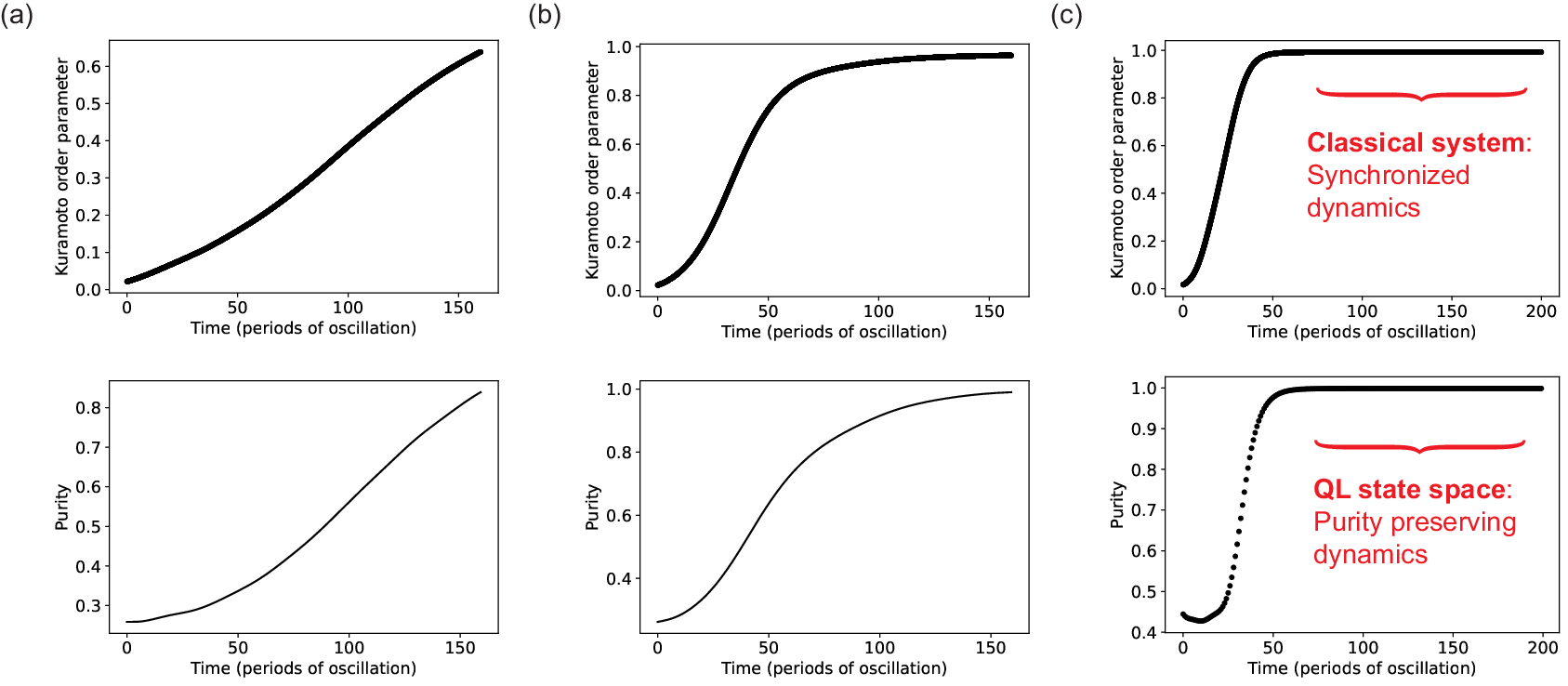}
	\caption{Ensemble-averaged Kuramoto order parameter as a function of time predicted for the system of oscillators and compared to the purity of the emergent QL state with largest eigenvalue. When the coupling, $K$, is sufficiently large, the system equilibrates to limit cycle dynamics, indicating a synchronized classical system. The state space concomitantly evolves to an equilibrium where purity of the emergent state is preserved. (a) $K = 100$, averaged over 100 iterations. (a) $K = 200$, averaged over 100 iterations. (a) $K = 400$, averaged over 500 iterations. }
		\label{figSync4}
\end{figure*}

We show the purity-preserving regime in Fig. \ref{figSync4}. In the figure, calculations with various parameters are compared, as specified in the caption. Each simulation was averaged over at least 100 state preparations. After an initial period when the oscillators synchronize, the system settles to equilibrium dynamics. When $K$ is sufficiently large the oscillators are locked in a limit cycle, that is, a closed, isolated (stable) phase trajectory. Owing to the way we define the oscillator network of each QL bit, these stable dynamics involve classical superpositions of oscillators local to each subgraph. In the corresponding state space, we find that when the classical system is synchronized, the purity of the emergent QL state is approximately 1. 

Phase coherence of the basis states is essential for sustaining stable superposition states. This phase coherence is controlled by the degree of synchronization of the classical network. Depending on the properties of the classical phase oscillator network, therefore, the QL states can exhibit purity-preserving evolution, or they can be completely mixed, or any case in between. When the oscillator network is synchronized, that is, the phases are coherent, then the QL states have high purity.  

\subsection{Dephasing}

In the previous section we studied how synchronization of the classical network produces robust QL states. We now consider a classical network where the couplings are weak compared to the standard deviation of frequency offsets. We thereby ensure that the system cannot synchronize. We set the state preparation so that all the oscillators are initially synchronized. That is, the von Mises distribution of oscillator phases is narrow. We hypothesize that, with these model parameters, a QL state that initially has high purity, will become more mixed with time as the network de-synchronizes.  That is, the opposite trend to that illustrated in Fig. \ref{figSync4} will ensue. 

An obvious example is a network of uncoupled oscillators with a frequency distribution. Starting in-phase, free evolution of the oscillators will dephase the order parameter and states in the network. The mechanism for such dephasing is similar to that for decoherence of a qubit\cite{Zurek2003}. Each coefficient in the chosen basis for the effective states acquires a phase offset from the mean as a function of time. Let's write this as $c_m^i = e^{i\zeta_mt}$. Then an off-diagonal density matrix element in this basis evolves as
\begin{equation}
	\rho^i_{mn}(t) = \sum_{\text{prep}} e^{i(\zeta_n - \zeta_m)t} .
\end{equation}
In the average, the phases in the emergent states evolve randomly because of the complexity of the underlying classical system, which leads to loss of phase coherence in these off-diagonal density matrix elements (they decay). This dephasing is independent of how we project the effective states because all the oscillators in the network are unsynchronized.

\begin{figure*}
	\includegraphics[width=12 cm]{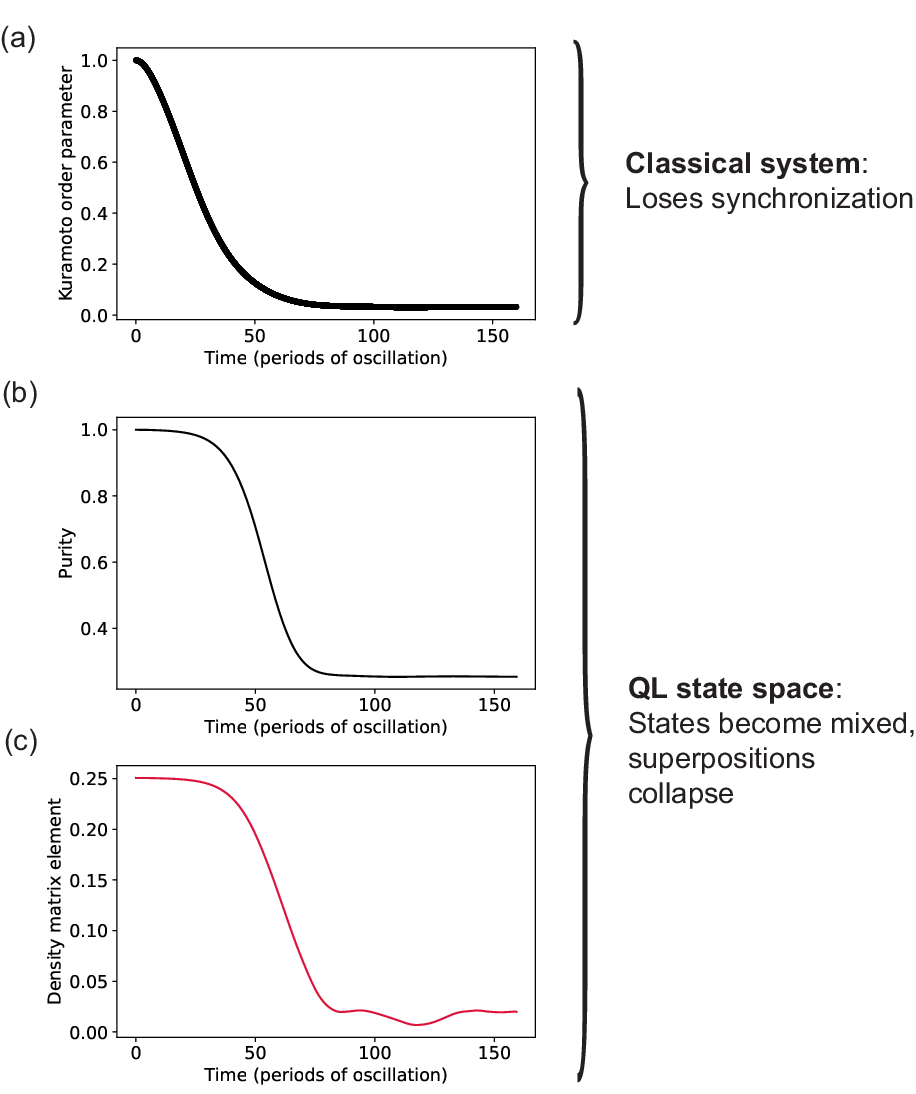}
	\caption{Results of calculations of an oscillator system comprising the Cartesian product of two QL bits, using the same procedures and parameters as the previous calculations, except here we set the coupling to be weak, $K = 30$ and the initial condition sets all the oscillators in phase. The averaging is carried out over a single graph.  (a) The order parameter, initially 1, indicating the initial synchronization condition, decays as the oscillators become out of phase. (b) As the oscillators dephase, the purity of the emergent state decays to an equilibrium mixed state. (c) Concomitantly, off-diagonal elements of the density matrix in the product basis decay to zero, highlighting loss of superpositions.}
		\label{figSync5}
\end{figure*}  

In Fig. \ref{figSync5} we show results of calculations of an oscillator system comprising the Cartesian product of two QL bits, using most of the same procedures and parameters as the previous calculations. Two key parameters were changed: the coupling was set to be weak, $K = 30$, and the initial condition sets all the oscillators in phase. The average is over 500 realizations of state preparation.

The order parameter is initially 1, indicating the initial synchronization condition, then it decays as the oscillators become out of phase. The dephasing is initiated by the fastest and slowest oscillators losing synchronization, which disconnects the network sufficiently so that, at a critical time in the evolution, all oscillators rapidly lose synchronization. This happens essentially by free evolution of the oscillators according to their natural frequencies because the coupling is too small to enforce synchronization. Accompanying the loss of synchronization, the purity of the mixed state decays. The fully mixed state has a purity of 0.25 because the density matrix is that for a four-state system\cite{ScholesQIS}.

The model captures the concept of an isolated QL system that succumbs to dephasing spontaneously. The numerical study shows an example of an initial pure state that decays to a final, equilibrium, state that is fully mixed in the product basis. The superpositions between basis states are completely lost, as evidenced by decay of the off-diagonal elements of the density matrix in the product basis, Fig. \ref{figSync5}c. The classical variables---the phase oscillator network---control the coherence of the state, such that the quantum state space is slaved to the dynamics of the network. When the network is poorly synchronized, its many degrees of freedom act as a kind of environment that decoheres the states\cite{Zurek2003}.  

\subsection{Purity preserving dynamics}

Despite the fact that the underlying synchronization dynamics are inherently nonlinear, maps in the state space taking the state from one time point to another, $\Xi: \rho(t_1) \rightarrow \rho(t_2)$, are linear maps. As discussed earlier in the paper, the reason for this linear representation of a nonlinear systems boils down to the way Eq. 12 evolves the edge biases in the adjacency matrix by a unitary operator.
\begin{Proposition}
	Maps of quantum-like (QL) states $\Xi: \rho(t_1) \rightarrow \rho(t_2)$ generated by a classical system of coupled phase oscillators are linear maps.
\end{Proposition}
To prove the proposition, first note that we construct each state $\rho$ as a convex combination of $\rho_j$ taken from trajectories indexed by $j$. Recall that the statistical averaging was done to account for an unknown initial phase distribution of the oscillators. So, given $M$ trajectories in the ensemble,
\begin{equation*}
	\frac{1}{M} \sum_j \rho_j(t_1) \rightarrow \frac{1}{M} \sum_j \rho_j(t_2),
\end{equation*}
which is obviously linear. It remains to consider maps within each trajectory that take the pure state $\rho_j(t_1)$ to $\rho_j(t_2)$. This amounts to examining maps of the coefficients in Eq.25. Consider the map $X: \Phi_j(t_1) \rightarrow \Phi_j(t_2) $, where $\Phi_j(t_k)$ is the matrix of oscillator phases, Eq. 13, for trajectory $j$ at time point $t_k$. $X$ simply takes the phases for the oscillators at time point $t_1$ and transforms them to the new phases at $t_2$, that is, changing each phase indexed by $i$ by $\Delta\theta_i = \theta_i(t_2) - \theta_i(t_1)$. The map is the matrix containing $e^{i\Delta\theta_i}$ at each diagonal entry $i$. Thus, $\Phi_j(t_2) = X\Phi_j(t_1)$, where $X$ is a linear map. Then, by using the facts that unitary maps are linear and compositions of linear maps are linear, the proposition is proved. 

The requirement for a \emph{dynamical} classical system that mimics an analogous quantum system is that the dynamics are unitary. Thus the QL system must be linear in the sense that for all emergent states $x$ and $y$ of the QL system, maps act as linear operators $T$,
\begin{eqnarray}
	T(x + y) = Tx + Ty \\
	T(\alpha x) = \alpha Tx .
\end{eqnarray}
This condition is fulfilled for QL systems represented as systems of phase oscillators, as just shown. However, these maps should also be time invariant\cite{Jordan}. Then the dynamics of the QL states are unitary. This latter condition is not generally found for classical systems, where nonlinear dynamics are common.  

Therefore, unitary evolution of the emergent state should be found as limiting case of dynamical classical systems where the nonlinearity vanishes. Let's say the dynamics of the oscillator system are prescribed by a set of coupled differential (nonlinear) equations in the form of the Kuramoto model. When the coupling $K$ is sufficiently large relative to the interplay of other parameters, then numerical results suggest the phase oscillators strongly synchronize and we propose that the states evolve unitarily. 

\begin{Proposition}
	Purity-preserving dynamics arises as a limit of the dynamical evolution of a nonlinear phase-oscillator network if and only if the phases of the oscillators are locked (i.e. constant).
\end{Proposition}

The proof in the forward direction is: When the oscillators are synchronized, the system is globally equilibrated to a fixed point, where each $\dot{\theta}_i = 0$, so that 
\begin{equation}
	\epsilon_i = \frac{K}{N} \sum_{j=1}^{N} a_{ij} \sin(\theta_j - \theta_i).
\end{equation}
Notice also that as $K$ becomes sufficiently large, then $\sin(\theta_j - \theta_i)$ becomes small, implying we approach the limit that $\theta_j \approx \theta_i$.

For the proof in the other direction, notice that if the phases are constant (locked), then $\sin(\theta_j - \theta_i)$ is constant and hence the dynamics are linear. 

\section{Summary of the main findings}

In recent work we described how a classical system can be designed to produce a state space that has many properties similar to a quantum state space\cite{ScholesQLstates, QLproducts}. These classical systems, that we call quantum-like (QL), can be identified with networks that have a particular structure. Here we studied the time evolution of the QL states that are associated with a QL network that evolves by nonlinear dynamics. In particular, we sought to understand how the nonlinearity of the classical system manifests in the linear representation of the state space. To that end, we studied how a classical network is viewed from the perspective of the time evolution of its emergent state. 

We found that the underlying classical system not only provides a mechanism for producing the QL state space, but, in addition, it actively controls properties of the states. Owing to the complexity of the classical system (the system of coupled phase oscillators), we could identify different kinds of evolution of the QL states. In a particular regime where the oscillators do not synchronize, because the couplings $K$ between the classical phase oscillators are small, the phase oscillator network serves as a kind of bath, endowing the QL system with intrinsic properties like those characteristic of an open quantum system. For example, the system may exhibit a form of  \emph{intrinsic} decoherence\cite{Stamp2012}. That is seen in Fig. \ref{figSync5}, where the mixedness of the emergent QL state \emph{increases} as the oscillators in the classical system \emph{de-synchronize}. In contrast, when $K$ is large enough, the classical network spontaneously synchronizes. Then, as shown in Fig. \ref{figSync4}, the mixedness of the emergent QL state \emph{decreases} as the oscillators in the classical system synchronize (the purity of the state increases).

\section{Appendix A. Synchronization}

To explain the  mechanism underlying synchronization, we review the Van der Pol method, using a derivation reproduced from Chapter 12 of ref \cite{Nekorkin2015}. Consider the quasilinear oscillator equation,
\begin{equation}
	\ddot x + \omega_0^2x = \mu f(x, \dot x)
\end{equation}
where the dots indicate differentiation with respect to time. We re-write the equation in the equivalent form
\begin{eqnarray}
	\dot x = y \\
	\dot y = -\omega_0^2x + \mu f(x, y)
\end{eqnarray}
where $0 < \mu << 1$ is a small parameter that determines how close the system is to a linear conservative system. Now define polar coordinates
\begin{eqnarray*}
	\rho = \sqrt{x^2 + y^2/\omega_0^2} \\
	\tan \phi = \omega_0x/y
\end{eqnarray*}
then find that Eqs 2, 3, with $\mu = 0$ reduce to the phase paths of the harmonic oscillator, that is, circles about the origin:
\begin{eqnarray}
	x = \rho \cos (\omega_0t + \phi) \\
	y = -\rho \omega_0 \sin (\omega_0t + \phi)
\end{eqnarray}

For $\mu \neq 0$, we apply the identity $\cos (A + B) = \cos A \cos B - \sin A \sin B$ and the product rule to obtain from Eqs 2--5
\begin{eqnarray}
	\dot \rho \cos(\theta) - \rho \sin(\theta) \dot \phi = 0  \\
	-\dot \rho \omega_0 \sin(\theta) - \rho \omega_0 \cos(\theta) \dot \phi = \mu f(x, y)
\end{eqnarray} 
where  $\theta = \omega_0t + \phi$, $x = \rho \cos \theta$, and $y = -\rho \omega_0 \sin \theta$. The solutions are written
\begin{eqnarray}
	\dot \rho = \frac{-\mu f(x,y) \sin \theta}{\omega_0}   \\
	\dot \theta = \omega_0 - \frac{\mu f(x,y) \cos \theta}{\rho \omega_0} 
\end{eqnarray} 

Typically $\rho$ is slowly varying compared to $\theta$. The nonlinearity can arise for various reasons, and is reflected in $f(x,y)$, case by case. For example, the pendulum clock is subjected to regular, short pushes that counteract dissipative losses of the pendulum. Electronic oscillator circuits incorporate a nonlinear element that, with the right conditions, enables the circuit to produce sustained oscillations because of feedback between the oscillatory system and the nonlinear element. 

In the case of a system of many interacting oscillators, the Kuramoto model\cite{daFonseca2018, Strogatz2000} attributes the nonlinearity to the form of coupling among the oscillators:
\begin{equation}
	\dot{\theta}_i   = \nu_i - \frac{K}{N} \sum_{j=1}^{N} a_{ij} \sin(\theta_j - \theta_i),
\end{equation}
where $\nu_i $ is the frequency of the oscillator at node $i$, $\theta_i$ is the oscillator phase defined, as we did above, in terms of accumulated phase and an offset $\theta_i(t) = \epsilon_it + \phi_i$. $K$ is the coupling value, $a_{ij}$ (with value 0 or 1) are taken from the adjacency matrix of the graph. The $a_{ij}$ represent the node connectivities in the network. The coupling is renormalized by the factor $1/N$ so that networks are size-intensive with respect to the lowest eigenvalue. Notice that the nonlinearity comes about because of the terms $f(\theta_j - \theta_i)$, where $f(\dots) = \sin(\dots)$. 

This remarkably simple system of equations predicts how oscillators collectively synchronize, according to the chosen parameters such as frequency disorder, coupling strength, and initial phases of the oscillators.  Feedback introduced by small coupling between pairs of oscillators causes oscillators to speed up or slow down during a period of oscillation---according to their natural frequency relative to the mean. Ultimately, the oscillator system equilibrates to a steady state where, for some initial conditions, phases remain locked in step. It is entropically favorable to minimize $\theta_{ij} = \theta_j - \theta_i$ because otherwise the system has to maintain a sequence of correlations among the evolving phases, decided by the sum rule for angles in an $n$-gon, $\sum \theta_{ij} = \pi(n-2)$. That influences the allowed phase differences. For example, consider three oscillators in the network ($i, j, k$), where we know $\theta_{ij}$ and $\theta_{ik}$, then we must have that $\theta_{kj} = \pi - \theta_{ij} - \theta_{ik}$. Thus many correlations are embedded in the coupling matrix. 

The propensity for a system of oscillators to synchronize depends on the form of the coupling matrix, that is, the structure of the graph where each oscillator is represented by a vertex and pairwise coupling is indicated by an edge\cite{Scholes2020, Townsend2023}. Synchronization also tends to be more robust when many oscillators are coupled, provided that the couplings are global. Notably, systems with only nearest-neighbor coupling synchronize poorly compared to more highly connected networks\cite{Townsend1, Scholes2020}. 

Collective synchronization in an oscillator ensemble comprising $N$ oscillators is quantified by the order parameter, where $\text{Re}(x)$ means the real part of $x$:
\begin{equation}
	\text{order parameter} = \text{Re} \Big( \frac{1}{N} \sum_{i=1}^N e^{i\theta_i} \Big) .
\end{equation}

A connection between this synchronization model and random matrix theory was established in ref \cite{ScholesAbsSync}. In that work we considered a network of phase oscillators defined by a graph and considered a Gaussian distribution of frequencies for the oscillators. It was noted that an ensemble of synchronization trajectories, each started with a random phase distribution for the oscillators, equilibrated to a phase distribution representative of the order parameter and indicative of the standard deviation of the frequency distribution. That is, for example, when the standard deviation of the frequency distribution is less than the coupling $K$, but not much less, then the equilibrated phases are clustered, with a certain circular standard deviation which can be related to the frequency distribution. It was thus proposed that the equilibrium properties of the phase oscillator ensemble can be predicted using a phase-oscillator version of random matrix theory. A relevant finding is that the states of a poorly synchronized phase oscillator network are more mixed than those of a synchronized network. We use a related method in the calculations of spectra and states described below.

We can also consider systems comprising an infinite number of oscillators. A continuum limit of the Kuramoto model can be developed in terms of oscillator phase densities\cite{Acebron2005, Strogatz1992} and Sec. 7 of Ref. \cite{Strogatz2000}.

\section{Appendix B. Graph products}

We can produce new graphs from existing graphs using a product operation\cite{GraphProducts}. This allows systematic propagation of a property possessed by the base graphs. For example, in the case of the Cartesian product of graphs $G$ and $H$, described below, the chromatic number of the product graph is that of $H$ or $G$ (whichever is the larger chromatic number). For the present work, we want to propagate the way the QL bit graph represents a two-state system. Graph products are defined starting with the vertex set of the Cartesian product of vertex sets,
\begin{equation*}
	V(G) \times V(H) \rightarrow \underbrace{ \{(u,x) \} }_{\text{set of ordered pairs}},
\end{equation*}
where the set of ordered pairs enumerates all pairs of vertices, one taken from $V(G)$ and one from $V(H)$. Edges are determined by rules governed by the type of product\cite{GraphProducts}. There are four common graph products: the direct product $\times$ (also called tensor product or Kronecker product), the Cartesian product $\Box$, the strong product $\boxtimes$, and the lexicographic product $G[H]$.  The Cartesian product of graphs is the product relevant for our purposes. It is defined here:

\begin{Definition}
	(Cartesian product of graphs) $G \Box H$ is defined on the Cartesian product of vertex sets, $V(G) \times V(H)$. Let $\{u, v, \dots\} \in V(G)$ and $\{x, y \dots\} \in V(H)$. Let $E(G)$ and $E(H)$ be the set of edges in $G$ and $H$ respectively. The edge set of the product graph $G \Box H$ is defined with respect to all edges in $G$ and all edges in $H$  as follows. We have an edge in $G \Box H$ from vertex $(u,x)$ to vertex $(v,y)$ when
	\begin{itemize}
		\item either there is an edge from $u$ to $v$ in $G$ and  $x = y$,
		\item or there is an edge from $x$ to $y$ in $H$ and  $u = v$.
	\end{itemize}
\end{Definition}

The  spectrum of the Cartesian product (see, for instance \cite{Barik2018}) is given by

\begin{Proposition}\label{eq:eig_prod}
	(Spectrum of a Cartesian product of graphs) Given
	\begin{enumerate}
		\item[] A graph $G$, for which its adjacency matrix $A_G$ has eigenvalues $\lambda_i$ and eigenvectors $X_i$, and
		\item[] A graph $H$, for which its adjacency matrix $A_H$ has eigenvalues $\mu_i$ and eigenvectors $Y_i$, then
	\end{enumerate}
	the spectrum of $G \Box H$ contains eigenvalues $\lambda_i + \mu_j$ and the corresponding eigenvectors are $X_i \otimes Y_j$.
\end{Proposition}

Let's consider an example of spectra of graph products based on a model cycle graph on five vertices ($C_5$). Spectra of products of the 5-cycle, $C_5$, are displayed in Fig. \ref{figSync3}a-c. The largest eigenvalue of $C_5$ is $\lambda_0(C_5) = 2$, and the second eigenvalue $\lambda_1(C_5) = 0.62$. Thus, for the products we find $\lambda_0(C_5 \Box C_5) = \lambda_0(C_5) + \lambda_0(C_5) = 4$ and $\lambda_1(C_5 \Box C_5) = \lambda_0(C_5) + \lambda_1(C_5) = 2.62$, and so on. Notice that the gap between the highest two eigenvalues, $\lambda_0 - \lambda_1$, remains constant as we take products, so that when we take products of expander graphs, emergent states in the base graphs are emergent states in the product. 

\begin{figure*}
	\includegraphics[width=13.5 cm]{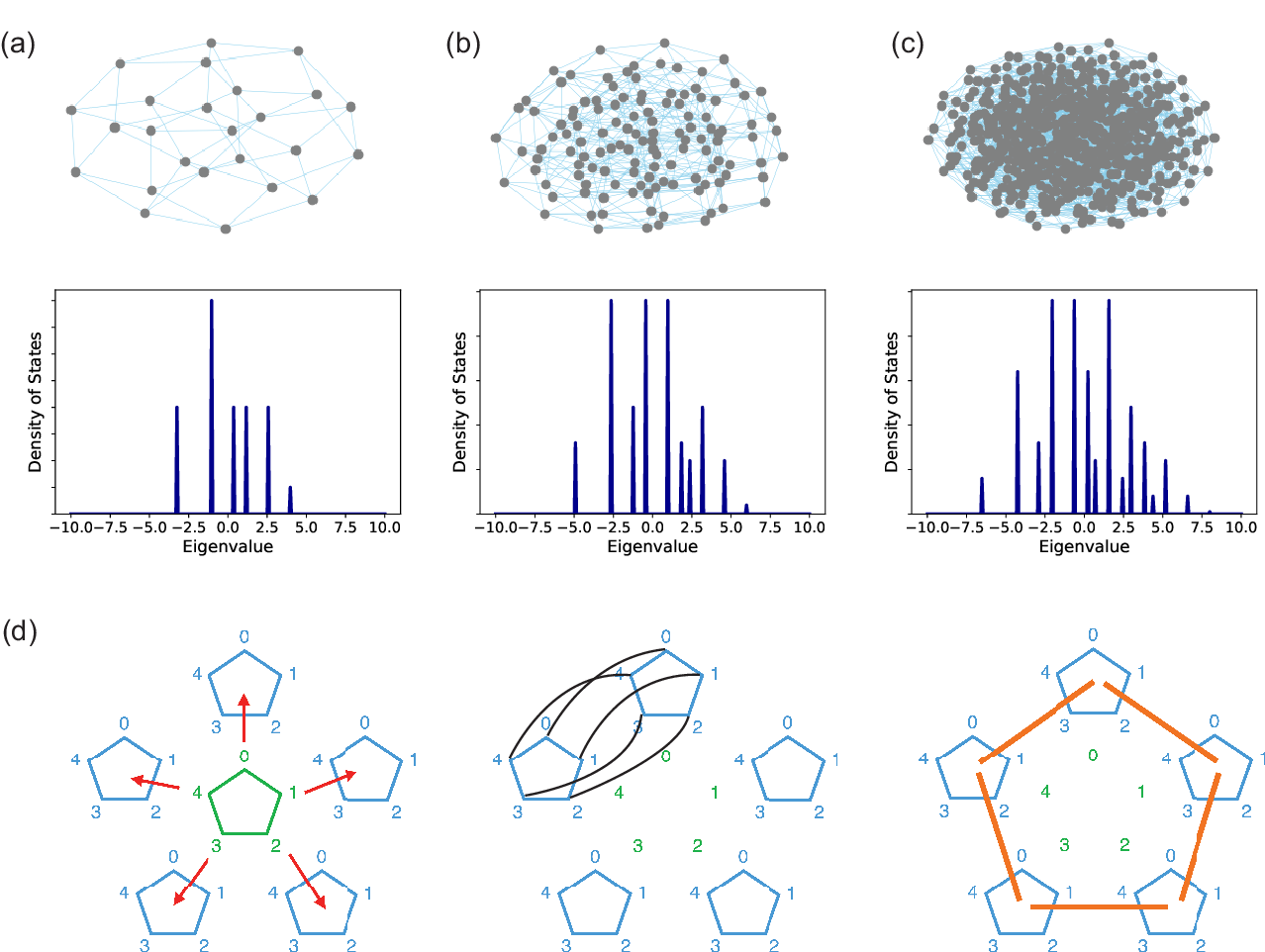}
	\caption{Examples of graph Cartesian products and corresponding spectra for (a) $C_5 \Box C_5$ (b) $C_5 \Box C_5 \Box C_5$ (c) $C_5 \Box C_5 \Box C_5 \Box C_5$. (d) Procedure for the physical construction of  the product $C_5 \Box C_5$, see text. The vertices of each base graph ($C_5$) are labeled $0, 1, \dots, 4$. The vertices of the product graph $G \Box H$ are, accordingly, $(i,j)$, where $i$ labels a vertex in $G$ and $j$ labels a vertex in $H$. }
	\label{figSync3}
\end{figure*}   

In Fig. \ref{figSync3}d we show how the product $C_5 \Box C_5 = G \Box H$ is produced explicitly. Let's label the graphs $G$ and $H$. One of the graphs, say $H$ (drawn in green), templates the product. For each vertex in the graph $H$ we draw one copy of the the graph $G$ (the blue graphs). The vertices are indexed by the index pair, one corresponding to the graph $G$ and one index deriving from the vertex of $H$ associated with each copy of $G$. We then use these second indices to draw edges between identical vertices of the copies of $G$  templated according to the edges in $H$. One set are shown as the black edges. Finally we have the product graph, which can be drawn to display explicitly the graphs $G$, as shown in the figure, or we could draw a similar picture that clearly shows five copies of $H$. The graph product thereby gives a physical picture of the tensor product.

Notice that the Cartesian graph product $G_A \Box G_B$ is constructed by putting a copy of $G_A$ at each vertex of $G_B$, then completing the additional edges. This mirrors the way a tensor product of $n$-dimensional Hilbert spaces $\mathcal{H}_A \otimes \mathcal{H}_B$ can be viewed as an $n$-fold direct sum of $\mathcal{H}_A$. See Remark 2.6.8 in ref \cite{KadRing1} for a proof. Also note that the method for drawing the product graph indicates how the corresponding adjacency matrix is structured in block form.

\section{Appendix C. The Lohe model for synchronization of quantum bits}
An explicit theory for synchronization of quantum bits was proposed by Lohe\cite{Lohe2009, Lohe2010, Choi2015}. The theory is developed by writing a generalized non-Abelian version of the Kuramoto model for coupled $n$-level systems:

\begin{equation}
	i\hbar \dot{U}_i U^\dagger_i  = H_i - \frac{iK}{2N} \sum_{j=1}^{N} a_{ij} (U_j U^\dagger_i - U_i U^\dagger_j ).
\end{equation}

Here, $H_i$ is a constant $n \times n$ Hermitian matrix with eigenvalues being the natural frequencies of each oscillator (that is, the mean frequency plus the offset from the mean). The $U_i$ are $n \times n$ complex unitary matrices. Setting $n = 1$ gives $U_i = e^{-i\theta_i}$, recovering the Kuramoto model. Setting

\begin{equation}
	U_i = 
	\begin{bmatrix}
		\alpha_i & -\beta_i \\
		\bar{\beta}_i & \bar{\alpha}_i 
	\end{bmatrix}
	=
	\begin{bmatrix}
		w_i + iz_i & y_i + ix_i \\
		-y_i + ix_i & w_i - iz_i
	\end{bmatrix}
\end{equation}
yields a theory for predicting synchronization dynamics of $N$ coupled two-level quantum systems (quantum bits). The matrix in Eq. 22 can be written as a real unit 4-vector $\mathbf{x}_i = (x_i, y_i, z_i, w_i)$, defining how each qubit vector lies on $\mathbb{S}^3$. The vectors evolve under the continuous group operations of $\mathsf{SU}(2)$. Unsynchronized quantum bits are seen as the $U_i$ dispersed around $\mathbb{S}^3$, whereas synchronized quantum bits lock into phase on the manifold.

The Lohe model can be considered to be a specialized version of the model we study here because, first, it locks the phases of vertices within each QL bit and, second, it applies to the single excitation subspace of the tensor product basis. In the Kuramoto model, we can imagine synchronization of phase oscillators to require a choreographed speeding up of low-frequency oscillators and slowing of high-frequency oscillators. This happens because of the way the interaction between all pairs of oscillators depends explicitly on their phase difference, via $\sin(\theta_j - \theta_i)$.  Owing to the way we have set up the graphs to represent two-state systems---QL bits---we can apply the Kuramoto model to the oscillators and effectively synchronize QL bits. That is, our graph construction of QL bits can simply be used with the Kuramoto model. Whereas if we choose to treat vertices as qubits, then we should apply the Lohe model. These two approaches become similar (in principle yielding identical predictions) when phases are strongly synchronized within each QL bit graph, but arbitrary between graphs. The phase difference enters in a more complicated way in the Lohe model for coupled quantum bits, as we now outline.

Lohe\cite{Lohe2010} writes the $H_i$ so as to indicate the way each oscillator traces out a periodic path on $\mathbb{S}^3$ (under the continuous group operations of $\mathsf{SU}(2)$):
\begin{equation}
	H_i = \sum_j \omega_i^j \sigma_j = 
	\begin{bmatrix}
		\omega_i^3 & \omega_i^1 - i\omega_i^2 \\
		\omega_i^1 + i\omega_i^2& -\omega_i^3
	\end{bmatrix}
\end{equation}
where $\omega_i^j$ are the frequencies of the vector in $\mathbb{S}^3$, $\omega_i^1$, $\omega_i^2$, and $\omega_i^3$. For intuition about what the three frequencies mean, recall that a vector moving on $\mathbb{S}^2$, the unit sphere embedded in 3-dimensional space, can be indicated by the frequencies it sweeps out over two angles, the azimuthal angle and the polar angle. Then, for the unit 3-sphere, in 4-dimensional space, we need three unique angles. 

Now, substituting Eqs. 22 and 23 into Eq. 21, Lohe obtained
\begin{equation}
	\dot{\mathbf{x}}_i = \Omega_i \mathbf{x}_i - \frac{K}{N} \sum_{j=1}^N a_{ij}  \big[ \mathbf{x}_j  - \mathbf{x}_i (\mathbf{x}_j  \cdot \mathbf{x}_i ) \big]
\end{equation}
where 
\begin{equation*}
	\Omega_i = \mathbf{\omega}_i  \cdot \mathbf{L} = \omega_i^1 L_1 +  \omega_i^2 L_2 +  \omega_i^3 L_3 ,
\end{equation*}
with the $\mathbf{L}$ being the $4 \times 4$ skew symmetric matrices\cite{Farebrother}:
\begin{eqnarray*}
	L_1 = 
	\begin{bmatrix}
		0 & 0 & 0 & -1 \\
		0 & 0 & -1 & 0 \\
		0 & 1 & 0 & 0 \\
		1 & 0 & 0 & 0
	\end{bmatrix} , \\ 
	L_2 = 
	\begin{bmatrix}
		0 & 0 & 1 & 0 \\
		0 & 0 & 0 & -1 \\
		-1 & 0 & 0 & 0 \\
		0 & 1 & 0 & 0
	\end{bmatrix} , \\ 
	L_3 = 
	\begin{bmatrix}
		0 & -1 & 0 & 0 \\
		1 & 0 & 0 & 0 \\
		0 & 0& 0 & -1\\
		0 & 0 & 1 & 0
	\end{bmatrix} .
\end{eqnarray*}

From Eq. 24 we can see that synchronization now is promoted by the nonlinearity induced by the pairwise relationships between unit 4-vectors $\mathbf{x}_j  - \mathbf{x}_i (\mathbf{x}_j  \cdot \mathbf{x}_i )$.

It seems that we can emulate this synchronization of quantum bits by building a system of QL bits, interacting via edges connecting vertices of one QL bit to another. For consistency with the Lohe model, we would require each QL bit to comprise strongly synchronized oscillators, and to be synchronized. The emergent state of each QL bit gives its frequency. We can assign a random initial `phase' as the real 4-vector vector in $\mathbb{S}^3$ that defines the orientation of the eigenvector of the synchronized emergent state.  The key point is that this vector, associated to $|\psi_i \rangle$ in Eq. 6 of Ref. \cite{Lohe2010}, must represent the emergent state vector as an element that sweeps through values in $\mathbb{C}^2$, and we have established this to be the case. 

The system of $N$ QL bits would be combined into a new graph $G = G_A \oplus G_B \oplus \dots \oplus G_N$. The subgraphs are connected by edges, weighted so that the inter-QL bit coupling is weaker than the coupling between vertices within a QL bit. The graphs $G_A, G_B, \dots$ are the vertices of the Lohe model. This system of oscillators, propagated according to the Kuramoto model, should mimic the Lohe model. To compare the results we should project out the effective 2-state emergent states of each QL bit to construct the vectors $\mathbf{x}_i $.

\begin{acknowledgments}
This research was funded by the National Science Foundation under Grant No. CHE-2537080 and the Gordon and Betty Moore Foundation through Grant GBMF7114. I thank Andrei Khrennikov for suggesting the idea of comparing dynamics of the network to dynamics in the state space.
\end{acknowledgments}


\vspace{6pt} 




\bibliography{Scholes_bib_Sept2025}

\end{document}